\documentclass[a4paper,aps,pra,twocolumn,showpacs,preprintnumbers,amsmath,amssymb]{revtex4}
\usepackage{bbm}
\usepackage{amsmath}
\usepackage{verbatim}
\usepackage{graphicx}
\bibliography{References}
\def\kp{k_{\parallel}}
\def\kperp{k_{\perp}}

\begin{document}

\title{Harnessing vacuum forces for quantum sensing of graphene motion}

\author{Christine A. Muschik$^{1}$, Simon Moulieras$^{1}$, Adrian Bachtold$^{1}$,\\ Frank H. L. Koppens$^{1}$, Maciej Lewenstein$^{1,2}$, and Darrick E. Chang$^{1}$}

\affiliation{
$^{1}$ ICFO-Institut de Ci\`{e}ncies Fot\`{o}niques,
Mediterranean Technology Park, 08860 Castelldefels (Barcelona), Spain.\\
$^{2}$ ICREA - Instituci\`{o} Catalana de Recerca I Estudis Avan\c cats, 08010 Barcelona, Spain.
}

\begin{abstract}
Position measurements at the quantum level are vital for many applications, but also challenging. Typically, methods based on optical phase shifts are used, but these methods are often weak and difficult to apply to many materials. An important example is graphene, which is an excellent mechanical resonator due to its small mass and an outstanding platform for nanotechnologies, but is largely transparent. Here, we present a novel detection scheme based upon the strong, dispersive vacuum interactions between a graphene sheet and a quantum emitter. In particular, the mechanical displacement causes strong changes in the vacuum-induced shifts of the transition frequency of the emitter, which can be read out via optical fields. We show that this enables strong quantum squeezing of the graphene position on time scales short compared to the mechanical period.
\end{abstract}

\pacs{42.50.-p, 42.50.Lc, 34.35.+a, 42.50.Dv}

\maketitle

Vacuum forces cause attraction between uncharged objects due to the modification of the zero-point energy in the intervening space~\cite{Casimir1948,Lamoreaux2005}. They become extremely strong at short distances, which is considered to be a major problem: for example, they lead to stiction and are commonly  believed to be "one of the most important reliability problems in micro-electromechanical systems"~\cite{Spengen}.
However, one can also envision that the strength of vacuum forces enables them to be exploited for applications. A spectacular but challenging example is to engineer repulsive Casimir forces for frictionless devices and levitation~\cite{Lamoreaux2005,Capasso2009}. Here, we present an application possible with current experimental capabilities and without the need to create repulsion.\\

\begin{figure}[h!]
 \includegraphics[width=\columnwidth]{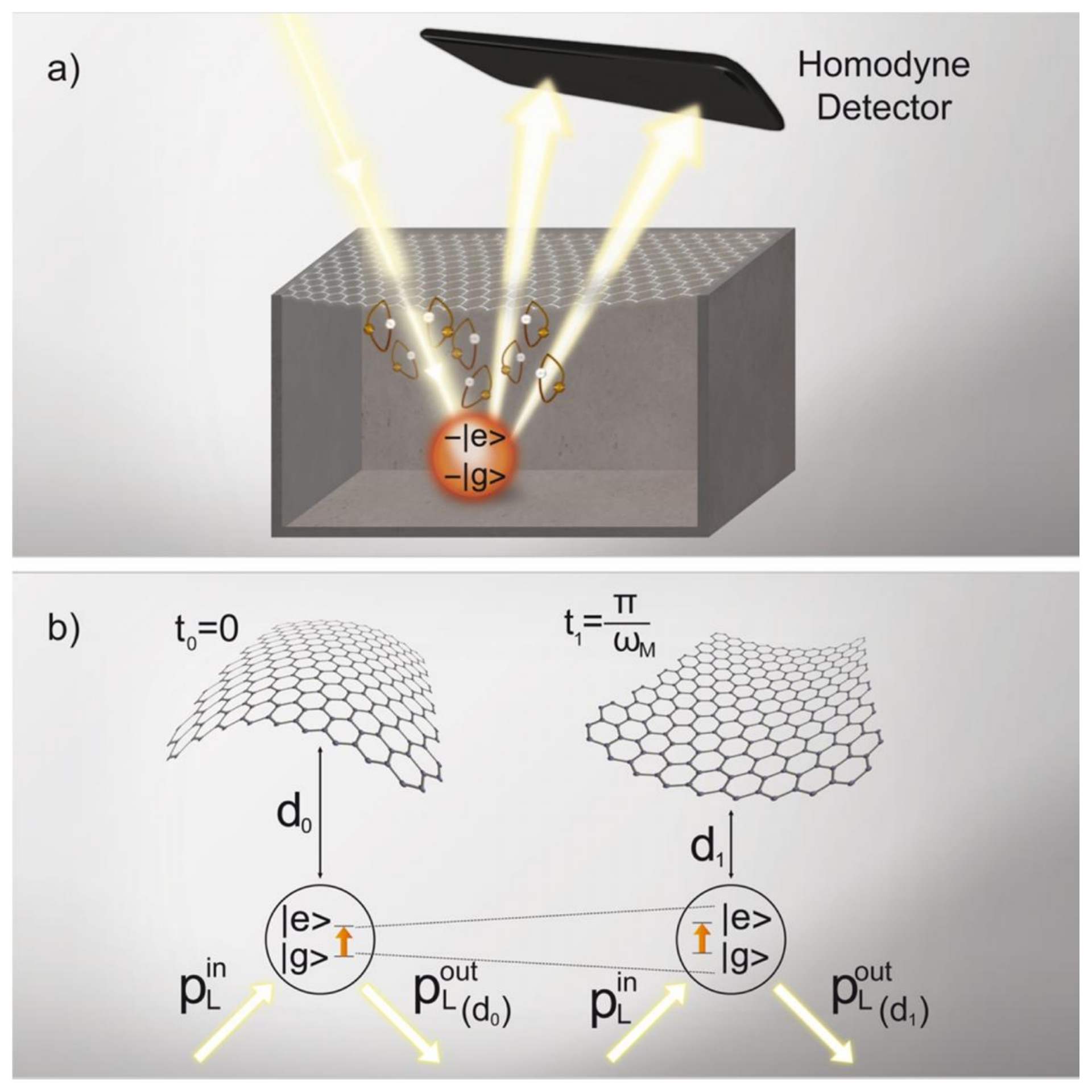}
 \caption{Motion sensing via vacuum potentials: a) A stationary emitter with states $|e \rangle$, $|g \rangle$ near a suspended graphene sheet is illuminated by a laser. Vacuum fluctuations~(illustrated by loops) affect the emitter's transition frequency, while the scattered light is measured by homodyne detection. b) The membrane and the quantum system interact via vacuum potentials: the emitter's energy levels are shifted depending on its distance $d$ to the membrane. The distance-dependent level shift translates into a phase shift of the light and can be read-out by measuring the $p$-quadrature of the scattered field.}
 \label{Fig:Setup}
 \end{figure}

We describe a technique that enables highly sensitive displacement detection of a mechanical system~\cite{Aspelmeyer2013}, which is critical for many devices such as force and mass sensors~\cite{Moser2013,Chaste2012}.
%
%
The ability to sense progressively smaller masses opens up new avenues for studying biological and chemical systems~\cite{Burg2007,Naik2009,Li2010,Grover2011} and finds exciting applications in surface science~\cite{Wang2010,Yang2011,Atalaya2011}.
A technological push towards faster high precision measurements would open up the possibility to observe a new class of phenomena paving the way towards the investigation of molecular diffusion processes and binding at the single molecule level.\\
%
%
\\Our scheme is based on the Casimir interaction between a surface and a quantum emitter: vacuum fluctuations lead to a modification of electronic state energies, which depends on the presence of nearby surfaces. A moving atom would therefore experience a force associated with the derivative of these shifts~\cite{Buhmann04,Buhmann07}. A stationary emitter experiences a measurable change in its resonance frequency that depends on the distance to the surface.
Finally, if the surface itself moves, such as the suspended nanomechanical membrane in Fig.~\ref{Fig:Setup}a, the modulation of the emitter's resonance frequency can be probed yielding an extremely sensitive displacement detection. This can be done by measuring the phase shift imparted on a field scattered by the emitter (Fig.~\ref{Fig:Setup}b).\\

There are several major advantages of our approach. First, vacuum interactions between an emitter and a surface are generic to any material. This provides a natural coupling to any mechanical element without the need to additionally functionalize or load it~\cite{Arcizet2011,Kolkowitz2012,Puller2012} or for the material to have low optical losses and high reflection (to integrate with an optomechanical system). Second, vacuum interactions are typically strong and divergent at short scales, providing a strong coupling between the mechanical system and the emitter.
We present a general formalism describing the detection of motion based on interactions with a nearby emitter. We describe realistic limits including back-action, emitter quenching, and imperfect measurement efficiency. We also analyze in detail the case where the mechanical system is a graphene resonator~\cite{Novoselov2004,McEuen2007,Hone2009}. This system is a particularly attractive candidate because its low mass and high Q-factor~\cite{Eichler2011} make it promising for a wide class of sensors. However, the capacitive coupling used in state-of-the-art detection techniques~\cite{Eichler2011,Gouttenoire2010,Yuehang2010} remains relatively weak. We show that it should be possible to generate a squeezed state of motion in a time short compared to the mechanical period, thus approximately achieving the limit of "projective measurement."\\

The Casimir potential for an emitter in its ground state at position $\mathbf{r}$ can be calculated~\cite{Buhmann04,Buhmann07} by considering its interaction with the vacuum modes of the electromagnetic field via the dipole Hamiltonian $H_{\text{\tiny{dip}}}=-\mathbf{d}\cdot \mathbf{E}(\mathbf{r})=-\sum_kg_k(\mathbf{r})(|e\rangle\langle g|+|g\rangle\langle e|)(a_k+a_k^{\dag})$, where $\mathbf{d}$ is the dipole moment of the emitter and $\mathbf{E}(\mathbf{r})$ is the electromagnetic field at position $\mathbf{r}$ with normal modes $k$. We consider a two-level system with states $|g\rangle$ and $|e\rangle$. $g_k$ denotes the vacuum Rabi coupling strength between the emitter and normal mode $k$ with creation operator $a_k^{\dag}$ and frequency $\omega_k$. The Casimir shift for an atom in its ground state arises from the non-excitation preserving terms of $H_{\text{\tiny{dip}}}$, which enables the ground state to couple virtually to the excited state and create a photon $|g,0\rangle\rightarrow|e,1_k\rangle$, which can be scattered from the surface before it is reabsorbed. The corresponding frequency shift of the ground state due to these fluctuations is given by $\delta \omega_g(\mathbf{r})=-\sum_kg_k(\mathbf{r})^2/(\omega_{0}+\omega_k)$, where $\omega_{0}$ is the unperturbed resonance frequency of the emitter. The shift can be re-expressed in terms of the classical dyadic electromagnetic Green's function $G(\mathbf{r},\mathbf{r},iu)$ evaluated at imaginary frequencies $\omega=i u$,
\begin{eqnarray}\label{Eq:LevelShift}
\delta\omega_{g}(\mathbf{r})=\frac{3 c\Gamma_0}{\omega_{0}^2}\int_0^\infty du\frac{u^2 }{\omega_{0}^2+u^2}\text{Tr} \{G(\mathbf{r},\mathbf{r},iu)\},
\end{eqnarray}
where $c$ is the speed of light and $\Gamma_0$ is the free-space emission rate of the excited state. Similar calculations allow one to determine the excited-state shift $\delta \omega_{e}$ and modified emission rate $\Gamma(\mathbf{r})$ near the surface~\cite{Prigogine1978,Schiefele2011}.
At distances $d$ much closer than the free-space resonant wavelength $\lambda_0$, the shift in the transition frequency of the emitter typically scales like $\Delta\omega=\delta\omega_e-\delta\omega_g \propto \Gamma_0/(d k_0)^3$ for a bulk material and like $\Delta\omega\propto \Gamma_0\alpha/(d k_0)^4$ for graphene, where $\alpha$ is the fine structure constant and $k_0=2\pi/\lambda_0$.\\
\begin{figure}
\includegraphics[width=\columnwidth]{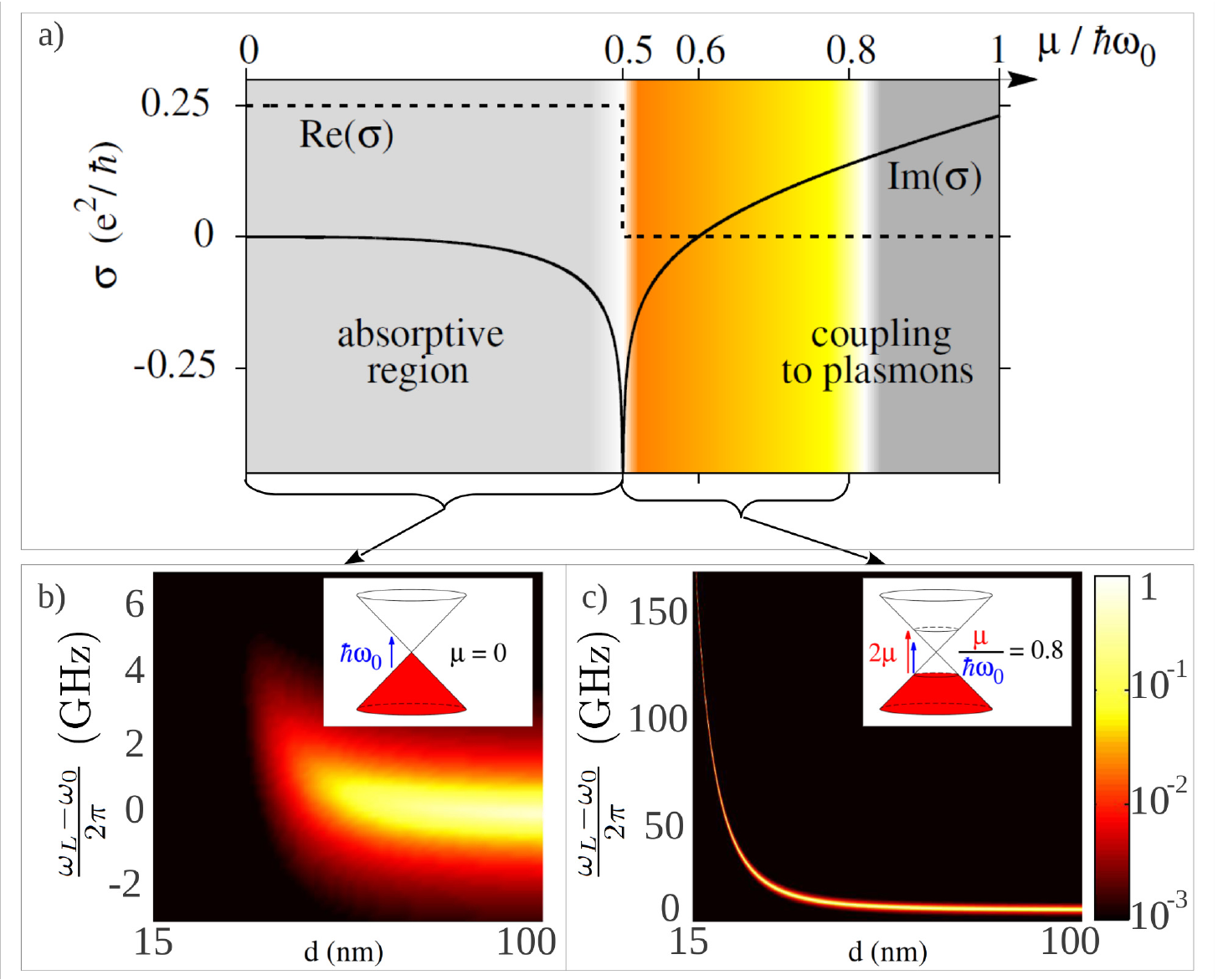}
\caption{Optical properties of graphene and frequency shifts on a nearby emitter: a) Conductivity $\sigma$ of graphene versus Fermi energy $\mu$ in units of $\hbar \omega_0$, where $\omega_0$ is the emitter's resonance frequency. The real part of the conductivity (dashed line) describes absorption. For $\frac{\mu}{\hbar \omega_0}<0.5$, light radiated by the emitter is absorbed since photons with frequency $\hbar\omega\geq 2\mu$ can induce inter-band transitions. For $\mu/\hbar\omega_0>0.6$, $\text{Im}\ \! \sigma>0$. This implies that the emitter can couple to surface plasmons, which increases its non-radiative decay rate;
b), c) Radiative photon scattering rate for low excitation power, $f(d,\omega_L)=\frac{\Gamma_{\footnotesize\textrm{rad}}(\Omega/2)^2}{(\Gamma/2)^2+(\omega_0+\Delta\omega-\omega_L)^2}$, normalized by the free-space resonant rate $f_0=\Omega^2/\Gamma_0$ for $\mu=0$ (b) and $\mu=0.8\hbar\omega_0$ (c). The shift~(broadening) of the peak versus distance reflects the emitter's frequency shift~(modified emission rate), while the decrease in contrast reflects increasing emission probability into non-radiative channels.
}\label{Fig:Conductivity}
\end{figure}

Here, we derive the sensing capability of a single mode of a mechanical system with a single emitter. Regardless of its complexity, any mechanical system can generally be decomposed into a set of normal modes with effective motional mass $m$, frequency $\omega_{\text{M}}$, displacement $x_{\text{M}}$ and momentum $p_{\text{M}}$~\cite{Safavi2013} and free Hamiltonian of any given mode
\begin{eqnarray}\label{Eq:HamiltonianM}
H_{\text{\tiny{M}}}=\!\frac{p_{\text{M}}^2}{2 m}\!+\!\frac{1}{2}m\omega^2_{\text{M}}x_{\text{M}}^2.
\end{eqnarray}
As previously described, the displacement of the mechanical system induces a position-dependent level shift on the emitter of the form $ H=\hbar\omega(x_{\text{M}}) \sigma_{z}$, where $\sigma_z =|e\rangle\langle e|-|g\rangle\langle g|$. As we are primarily interested in detecting small displacements, it is suitable to linearize
\begin{eqnarray*}
H_{\text{\tiny{int}}}=\hbar\omega(x_{\text{M}}) \sigma_{z}= \hbar g x_{\text{M}}\sigma_{z} +\mathcal{O}(x^2_{\text{M}}).
\end{eqnarray*}
The coupling coefficient $g=\frac{\partial\Delta\omega}{\partial x_M}$
describes the rate of change of the emitter frequency per unit displacement.\\

Next, we provide a quantum description of the emitter interacting with an external laser which probes the emitter's changing resonance frequency. This description consists of two parts, the dynamics of the emitter due to the incoming field, and the information about the emitter that is written onto the scattered light.
For the former, we restrict ourselves to the interaction with a laser field with Rabi frequency $\Omega$ and detuning $\Delta$ from the atomic transition at $x_{\text{M}}=0$, with Hamiltonian $H_{\text{\tiny{emitter}}}=\frac{\Delta}{2}\left(|e\rangle\langle e|-|g\rangle\langle g|\right)+\frac{\Omega}{2}\left(|e\rangle\langle g|+|g\rangle\langle e|\right)$. The latter is described by $a_{\text{L}}^{\text{out}}=-a_{\text{L}}^{\text{in}}+\sqrt{\nu\Gamma}\ \! |g\rangle\langle e|$, which relates the scattered fields to the atomic coherence. $\nu$ characterizes the detection efficiency, $\Gamma$ is the emitter's total~(surface-modified) emission rate, and $a^{\text{in}(\text{out})}$ is the annihilation operator of the light before (after) the interaction.\\

We consider the weak-driving limit, where the population of the atomic excited state is negligible. This limit is characterized by $\epsilon=\frac{\Omega^2}{\frac{\Gamma^2}{4}+\Delta^2}\ll 1$. Physically, working in the limit of $\epsilon\ll 1$ enables the emitter's dynamics to be linearized and ensures that the optical scattering is predominantly coherent.
Adiabatic elimination of the emitter yields an emitter-mediated interaction between the membrane and the light. The latter is described by its quadratures $x_{\text{L}}=(a_L+a_L^{\dag})/\sqrt{2}$, $p_{\text{L}}=-i(a_L-a_L^{\dag})/\sqrt{2}$. As explained in the Supplemental Material (SM)~\cite{SupplementalMaterial}, the reduced system evolves under the Hamiltonian $H=H_{\text{\tiny{M}}}\!+\!H_{\text{\tiny{ML}}}$, which contains a part describing free motion (M) (see Eq.~(\ref{Eq:HamiltonianM})) and a part describing the interaction between motion and light (ML),
\begin{eqnarray}\label{Eq:HamiltonianQND}
H_{\text{\tiny{ML}}}=\!\hbar\kappa\ \! x_M p_{\text{L}}.
\end{eqnarray}
The coupling constant $\kappa$ reflects the rate at which information about $x_{\text{M}}$ can be obtained and depends on the excited state population, coupling strength $g$, and detection efficiency $\nu$. It is given by $\kappa=2\bar{g}\sqrt{\epsilon\nu/\Gamma}$, where $\bar{g}=g\sqrt{2}\left(1-\frac{3}{8}\epsilon\right)$ is a renormalized coupling coefficient.
In the case of ideal detection efficiency, the rate at which information about $x_{\text{M}}$ (in vacuum units $x_{\text{\tiny{ZPM}}}=\sqrt{\hbar/(m\omega_M)}$) can be collected is given by $\kappa_{\text{\tiny{ideal}}}^2=\frac{4\epsilon\bar{g}^2}{\Gamma}$. Since we aim at measuring the mechanical motion on time scales which are short compared to $\omega_{\text{M}}^{-1}$, the dimensionless quantity $\frac{\kappa_{\text{ideal}}^2 x_{\text{\tiny{ZPM}}}^2}{\omega_{\text{M}}}=\frac{4\epsilon(\bar{g}\cdot x_{\text{\tiny{ZPM}}})^2}{\Gamma\omega_M}$ represents an important figure of merit characterizing the measurement strength.\\

The working principle of the scheme can be understood by considering the dynamics in the absence of undesired processes (which will be addressed below) during a short measurement time window $\Delta t\ll\omega_{\text{M}}^{-1}$. In this case, $H_{\text{\tiny{ML}}}$ leads to an evolution
\begin{eqnarray}
x^{\text{out}}_{L}&=&x^{\text{in}}_{L}+\kappa \sqrt{\Delta t}\ \! x_M^{in},\label{Eq:Mapping}
\end{eqnarray}
where the superscripts "in" and "out" denote operators before and after the interaction. For large $\kappa\sqrt{\Delta t}$, all motional properties are mapped onto the output field. Eq.~(\ref{Eq:HamiltonianQND}) also implies that the light imparts back-action onto the membrane, $p_{\text{M}}^{\text{out}}=p_{\text{M}}^{\text{in}}+\kappa\sqrt{\Delta t}\ \!x_{\text{L}}^{\text{in}}$, which affects the measurement precision for longer times $\Delta t > \omega_{\text{M}}$.\\
\begin{figure}
\includegraphics[width=\columnwidth]{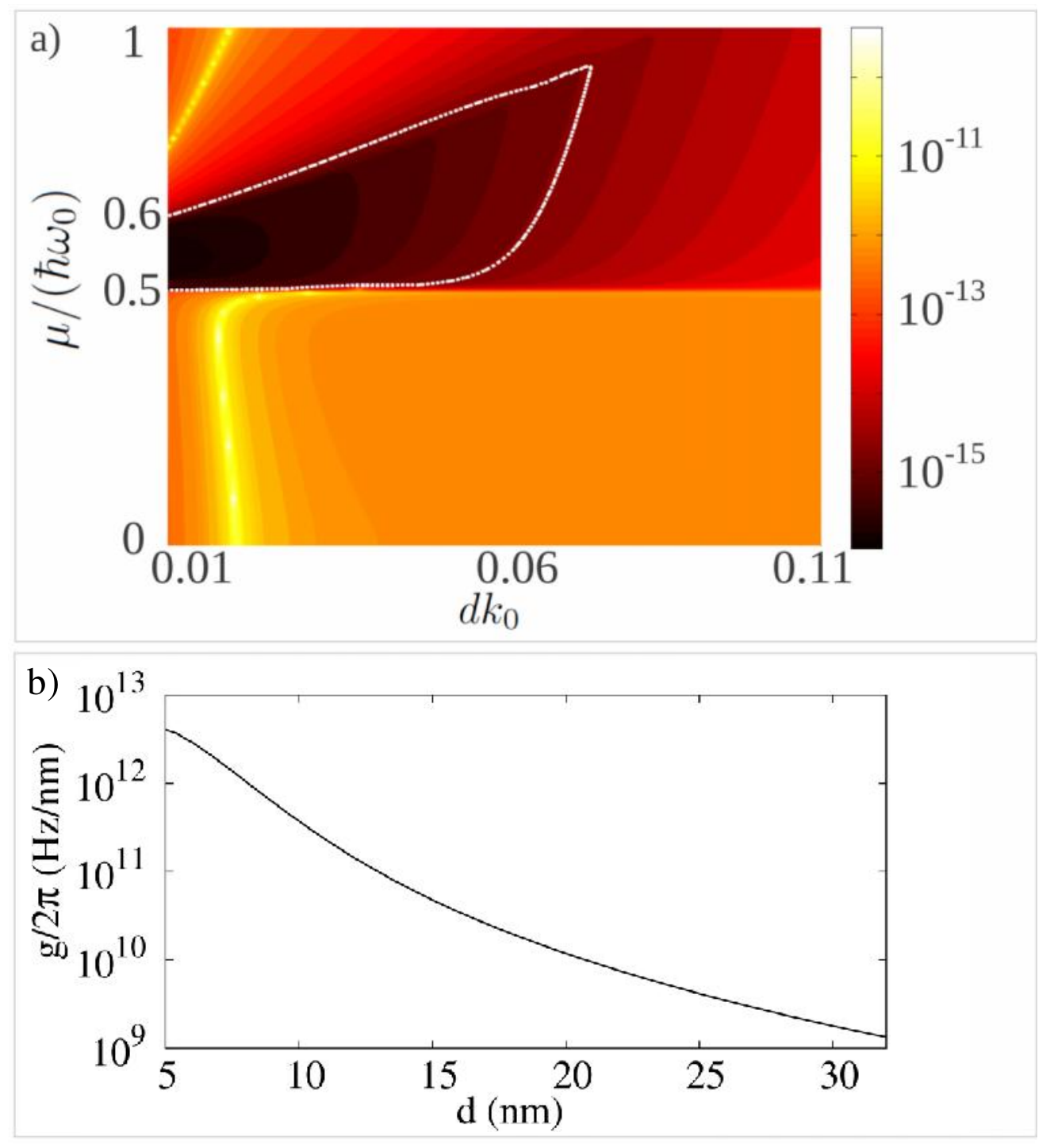}
\caption{Vacuum coupling and detection sensitivity:
a) Inverse coupling strength $\kappa^{-1}$ in $\frac{\text{m}}{\sqrt{\text{Hz}}}$ versus the distance $d$ and the Fermi energy $\mu$ in units of $\hbar\omega_0$, where $\omega_0$ is the emitter's resonance frequency in free space. Within the white line, $\kappa^{-1}<\frac{x_{\text{ZPM}}}{\sqrt{\omega_{\text{M}}}}$, indicating that measurements at the quantum level are possible with measurement times shorter than the oscillation period;
b) Coupling coefficient $g=\frac{\partial \omega}{\partial d}$ in $\frac{\text{Hz}}{\text{m}}$ versus the distance $d$ in nm for the Fermi energy $\mu=0.8\hbar\omega_0$ that yields the optimal coupling $\kappa$ for the considered parameters (see Fig.~4).
}\label{Fig:DynamicalProperties}
\end{figure}

While our analysis thus far was completely general, we now consider the case of graphene~\cite{Ribeiro2013,Scheel2013,Buhmann2010,Scheel2011,Klimchitskaya2014,Chaichian2012,Klimchitskaya2008a,Klimchitskaya2008b}, which has two complicating features. First, its "refractive index" (or more specifically, its conductivity) can be electrostatically tuned, which alters the level shifts through the Green's function in Eq.~(\ref{Eq:LevelShift}). Second, graphene can strongly "quench" or absorb light scattered by the emitter, yielding a fundamental upper limit on the detection efficiency $\nu$.
Here, we briefly summarize how these properties affect the overall sensitivity of our scheme (see SM~\cite{SupplementalMaterial} for details). Unlike in typical metals, the Fermi energy $\mu$ and associated conductivity $\sigma(\omega)$~\cite{Stauber2008} can be greatly tuned in graphene by applying a voltage~\cite{Novoselov2004} or by chemical doping and intercalation~\cite{Khrapach2012}. The conductivity directly influences how a proximal emitter interacts with the graphene leading to three different regimes as illustrated in Fig~2.
In the first regime of low Fermi level, $\mu<0.5\hbar \omega_0$, the conductivity is mostly real. Graphene is absorptive, as light can induce inter-band electronic transitions. The total emission rate of the emitter $\Gamma=\Gamma_{\text{rad}}+\Gamma_{\text{non-rad}}$  separates into radiative (i.e., free-space) and absorptive channels, with the latter dominating at close distances. Significant level shifts are observable, but with decreased free-space fluorescence (Fig.~2b).
The second regime of intermediate Fermi level yields optimal read-out sensitivity, as inter-band absorption becomes suppressed leading to a sharp decrease in $\text{Re}\ \! \sigma(\omega)$, while the level shift is maximum (Fig.~2c). In the third regime of high Fermi Level, $\mu\gtrsim 0.6\hbar \omega_0$, $\sigma(\omega)$ becomes mostly imaginary and positive, analogous to a thin conducting film. Such thin films support highly localized guided surface plasmons. The emitter can efficiently couple to these modes, which are dark to free-space detection channels and again result in a large $\Gamma_{\text{non-rad}}$~\cite{Koppens2011,Nikitin2011,Gaudreau2013}.\\

The implications can be seen in Fig.~3a, where we plot the sensitivity $\kappa^{-1}$ versus $\mu$ and the distance $d$. Non-radiative emission affects the sensitivity of our scheme since the detection rate $\kappa^2$ is proportional to the maximum possible detection efficiency $\nu$. In particular, $\nu=\eta_{\text{\tiny{det}}}\frac{\Gamma_{\text{\tiny{rad}}}}{\Gamma}$ contains one term $\eta_{\text{\tiny{det}}}$ describing the efficiency at which photons scattered in free space can be collected, and is technical in nature. The other term, $\frac{\Gamma_{\text{\tiny{rad}}}}{\Gamma}$, describes the probability for a photon to be scattered to free space~(versus absorbed by the material).
At an operating distance of $d=18$nm, the ideal Fermi level is $\mu=0.8$. As concrete example, we consider here $\epsilon=0.3$, $\eta_{\text{\tiny{det}}}=0.75$, $\Gamma_0=2\pi\times 240$~MHz and $\lambda_0=2\cdot10^{-6}$m and a graphene sheet with resonance frequency $\omega_M=2\pi\times1$~MHz and mass $m=2.81\cdot10^{-18}$kg. For these parameters, $\Gamma_{\text{rad}}/\Gamma=0.54$. The frequency shift per unit length is $g=2\pi\times 16\frac{\text{GHz}}{\text{nm}}$ (Fig.~3b), which compares very favorably to the best demonstrated couplings in cavity opto-mechanics experiments~\cite{Painter2010} and gives rise to a sensitivity of $\kappa^{-1}=5.6\cdot 10^{-16}\frac{\text{m}}{\sqrt{\text{Hz}}}$.\\
\begin{figure}
\includegraphics[width=\columnwidth]{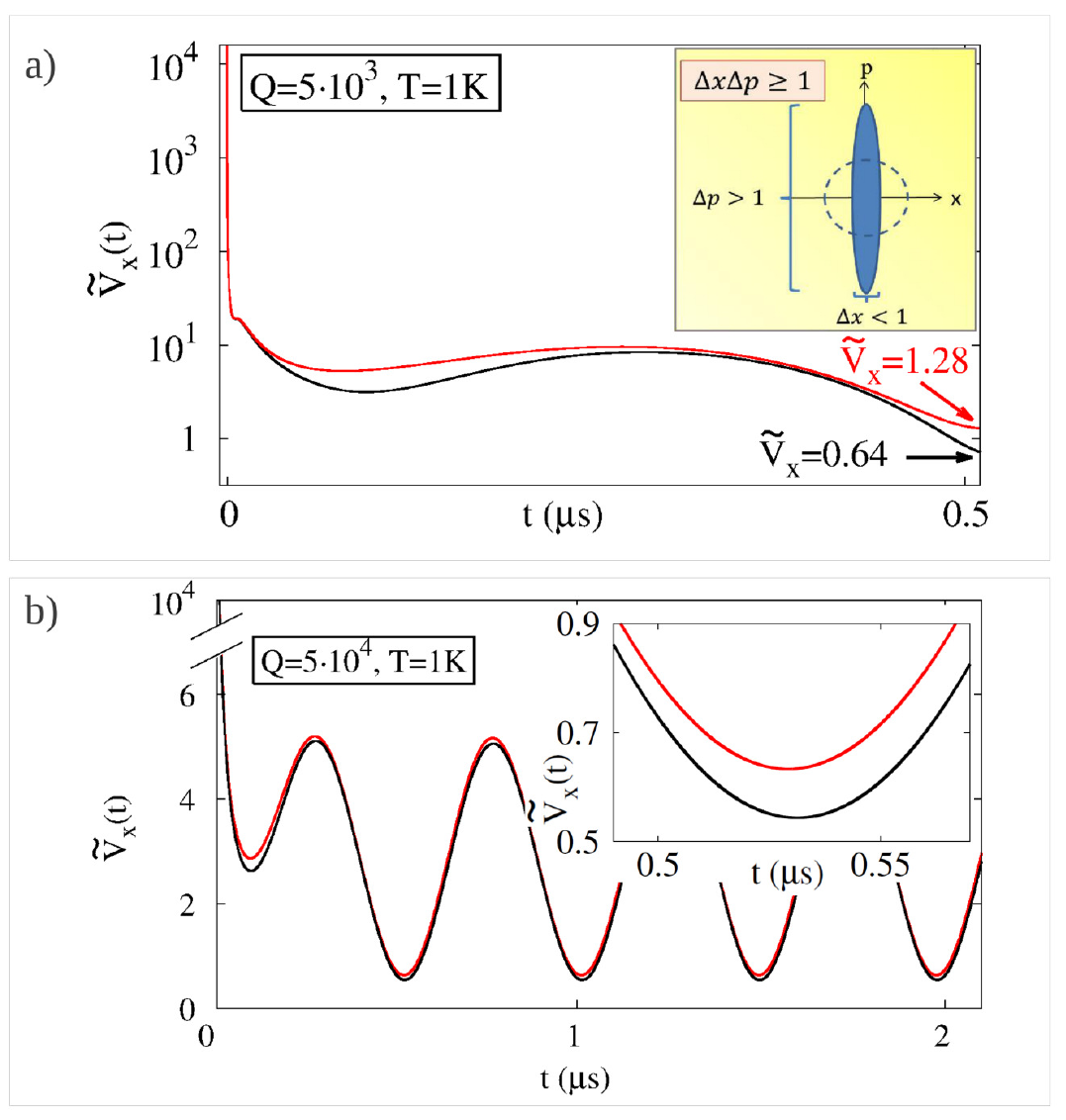}
\caption{Position squeezing of a thermal state at $T=1$K: The conditional variance of the position in the rotating frame, $\tilde{V}_x(t)$, is shown in units of the zero point fluctuations $x^2_{\text{\tiny{ZPM}}}$ versus time~(in $\mu$s) during a short time window for $Q=5\times10^3$ (a) and during several oscillation periods for $Q=5\times10^4$ (b). $\tilde{V}_x(t)<1$ certifies squeezing. Red (black) lines correspond to pure momentum damping $\gamma_x=0,\gamma_p=\gamma$ (symmetric damping $\gamma_x=\gamma_p=\gamma/2$).
The used parameters take the values $\omega_M=2\pi\times 1$~MHz, $m=2.8\times 10^{-18}$kg, $\mu=0.8\hbar\omega_0$, $d=18$nm, $\Gamma_0=2\pi\times240$~MHz, $\lambda_0=2\times 10^{-6}$m, $\eta_{\text{det}}=0.75$, $\epsilon=\frac{\Omega^2}{\Delta^2+\Gamma^2}=0.3$ (see main text).}\label{Fig:DynamicalProperties}
\end{figure}

The ability to perform highly sensitive position measurements on time scales that are short compared to $\omega_{\text{M}}^{-1}$ allows one create a squeezed state where the variance of the position of the graphene sheet $V_x=\langle x_{\text{M}}^2\rangle-\langle x_{\text{M}}\rangle^2$ is reduced below its zero-temperature variance $V_x<x_{\text{ZPM}}^2$. This comes at the expense of an increased variance of the momentum $V_p$ in compliance with the Heisenberg uncertainty principle $V_x\cdot V_p \geq\hbar^2/4$ (see inset Fig.~4a). The rotation in phase space would prevent the squeezing of $x_{M}$ or $p_{M}$, if measurements over several oscillation periods were required, but since the high coupling strength $\kappa$ allows for a fast and precise read-out, the Casimir scheme yields significant squeezing for realistic Q-factors~\cite{Chen2009}, as shown in Fig.~4. Similar results can be obtained for higher temperatures $T$ if the ratio $Q/T$ is kept constant.
The ability to perform fast position measurements is interesting for a number of reasons. For example, it can shed light on microscopic origins governing dissipation. As an example, we consider two different damping types: a symmetric model, where position and momentum are damped with equal rates $\gamma_x=\gamma_p=\gamma/2$, and pure momentum damping $\gamma_x=0$, $\gamma_p=\gamma$. In the latter case, almost noise-free position measurements can be made in the short time limit, \textit{i.e.}, if the measurement time $\Delta t$ is short compared to the rotation period in phase space, since momentum damping requires a time span on the order of $\omega_M^{-1}$ to affect the position. The high sensitivity of the scheme renders the distinction between different types of damping possible. Symmetric and momentum damping would become indistinguishable if averaged over several oscillation periods, but lead to different results if a high temporal resolution is available, as shown in Fig.~4a. An even greater degree of squeezing can be achieved if the incident light is modulated in time or if short pulses are used~\cite{Vanner2011}.\\

We have shown that quantum vacuum interactions can be a valuable resource for sensing at the quantum level. We have specifically analyzed the scheme for graphene, which is a promising platform for devices but currently lacks the means for fast readout. However, in principle, the presented method is quite general and applicable to a wide class of materials. If the separation between the membrane and the emitter is known, our scheme allows for the precise study and accurate measurement of Casimir forces~\cite{Sukenik1993,Landragin1996,Capasso2007,Bender2010,Alton2011}, which is an important step towards the vision of controlling and manipulating vacuum potentials. Finally, using specially engineered nanophotonic interfaces could provide even larger dispersive interactions in our scheme, which could lead to the generation of non-Gaussian quantum states of motion.
\begin{acknowledgements}
We gratefully acknowledge discussions with Joel Moser. This work was supported by the ERC grants QUAGATUA, CARBONLIGHT, CARBONNEMS and OSYRIS, the Alexander von Humboldt Foundation, TOQATA (FIS2008-00784), Fundaci\'o Privada Cellex Barcelona, the EU integrated project SIQS and the European Comission under Graphene Flagship (contract no. CNECT-ICT-604391).

\end{acknowledgements}

\appendix
\renewcommand{\figurename}{Supplementary figure}
\setcounter{figure}{0} \setcounter{equation}{0}
\renewcommand{\thefigure}{S.\arabic{figure}}
\renewcommand{\theequation}{S.\arabic{equation}}
\widetext
\section*{Supplemental Material}

In the following, we provide details of the Casimir sensing scheme presented in the main text. In Sec.~\ref{Sec:Casimir}, we address the effect of quantum vacuum potentials on a quantum emitter and explain how the modified energy level shifts and decay rates of a two level system close to a surface can be calculated. In Sec.~\ref{Sec:Linearization}, we derive the effective (linear) Casimir coupling Hamiltonian in the weak driving limit and in in Sec~\ref{Sec:Protocol}, we provide a detailed description of the optical read-out of the Casimir-potential induced level shifts and show how the motion of a membrane can be monitored using coherent light. Throughout the Supplemental Material, we use $\hbar=1$.
\section{Casimir-effect for quantum emitters}\label{Sec:Casimir}
The high sensitivity of the proposed sensing scheme is due to the large energy shifts that vacuum forces can induce in a quantum emitter close to a dielectric surface. In the following, we explain how these level shifts can be calculated.
The general expressions for the ground and excited state shifts $\delta\omega_{g}$ and $\delta\omega_{e}$ of an effective isotropic two-level emitter are given by~\cite{Buhmann}
\begin{eqnarray*}
\delta\omega_{g}(\mathbf{r})=\frac{3 c\Gamma_0}{\omega_{eg}^2}\int_0^\infty du\frac{u^2 }{\omega_{eg}^2+u^2}\text{Tr}\, G(\mathbf{r},\mathbf{r},iu),
\end{eqnarray*}
\begin{eqnarray*}
\delta\omega_{e}(\mathbf{r})=-\delta\omega_g(\mathbf{r})-\frac{\Gamma_0\pi c}{\omega_{eg}}\text{Tr}\,\text{Re}\,G(\mathbf{r},\mathbf{r},\omega_{eg}),
\end{eqnarray*}
where $G(\mathbf{r},\mathbf{r},\omega_{eg})$ is the classical (dyadic) electromagnetic Green's function. As described in the main text, the ground-state shift arises from excitation non-conserving terms in the atom-field interaction Hamiltonian involving the virtual emission and re-absorption of a photon. This contribution is most easily evaluated by rotating the arising integral to complex frequencies $\omega=iu$. The excited-state shift contains one term~($-\delta\omega_g(\mathbf{r})$) that arises from virtual emission and re-absorption of off-resonant photons from the excited state, and an additional term coming from the emission of a real photon~(proportional to the Green's function at the resonance frequency $\omega_{eg}$). Similarly, the spontaneous emission rate of this real photon can be modified in the presence of a dielectric surface~\cite{Buhmann},
\begin{eqnarray*}
\Gamma(\mathbf{r})=\Gamma_0+\frac{2\Gamma_0\pi c}{\omega_{eg}}\text{Tr}\,\text{Im}\,G(\mathbf{r},\mathbf{r},\omega_{eg}).
\end{eqnarray*}
The Green's function satisfies the equation
\begin{eqnarray*}
\left[(\nabla\times\nabla\times)-\frac{\omega^2}{c^2}\epsilon(\mathbf{r},\omega)\right]G(\mathbf{r},\mathbf{r}',\omega)=\delta(\mathbf{r}-\mathbf{r}')\otimes I.
\end{eqnarray*}
We approximate the Green's function of a suspended graphene mechanical resonator by that of an infinite graphene sheet, as the latter has an exact solution. This approximation is well-justified given that the regime of interest is one where the emitter sits at much closer distances $d$ to the graphene than the lateral size of the sheet $L$, $d\ll L$.

To be specific, we consider an infinite interface between vacuum and a dielectric surface located at $z=0$. The Green's function generally consists of an unimportant free term and a reflected component, the latter of which gives rise to position dependence in the level shifts and decay rates. Physically, this term describes the interaction of the emitter with its own field reflected from the surface.
For distances $z>0$~(on the vacuum side), the trace of this reflected component is
\begin{eqnarray*}
\text{Tr}\,G(z,z,\omega)=\frac{ic^2}{4\pi\omega^2}\int_0^{\infty}d\kp\,\frac{\kp}{\kperp}e^{2i\kperp z}\left(\frac{\omega^2}{c^2}r_s+(\kp^2-\kperp^2)r_p\right).
\end{eqnarray*}
Here $\kp$ and $\kperp$ are the parallel and perpendicular wavevector components, with $\kperp=\sqrt{(\omega/c)^2-\kp^2}$. The Fresnel reflection coefficients for $s$ and $p$-polarized waves in the case of graphene are given by $r_p=\frac{\kperp\sigma}{\kperp\sigma+2\epsilon_0\omega}$ and $r_s=-\frac{\mu_0\sigma\omega}{2\kperp+\mu_0\sigma\omega}$ and depend on the conductivity $\sigma$~\cite{Falkovsky2008}. The conductivity of graphene~\cite{Wunsch2006,Stauber2008} is given by
\begin{eqnarray}\label{Eq:Conductivity}
\sigma(\omega)=\frac{e^2\mu}{\pi}\frac{i}{\omega+i\gamma_g}+\frac{e^2}{4}\left[\Theta(\omega-2\mu)+\frac{i}{\pi}\log\left|\frac{\omega-2\mu}{\omega+2\mu}\right|\right],
\end{eqnarray}
where $\mu$ is the Fermi energy and $\gamma_g$ is a phenomenological parameter characterizing intraband losses. For our numerical simulations, we use $\omega_0/\gamma_g=10^{3}$.\\

\begin{figure}
 \includegraphics[width=0.35\columnwidth]{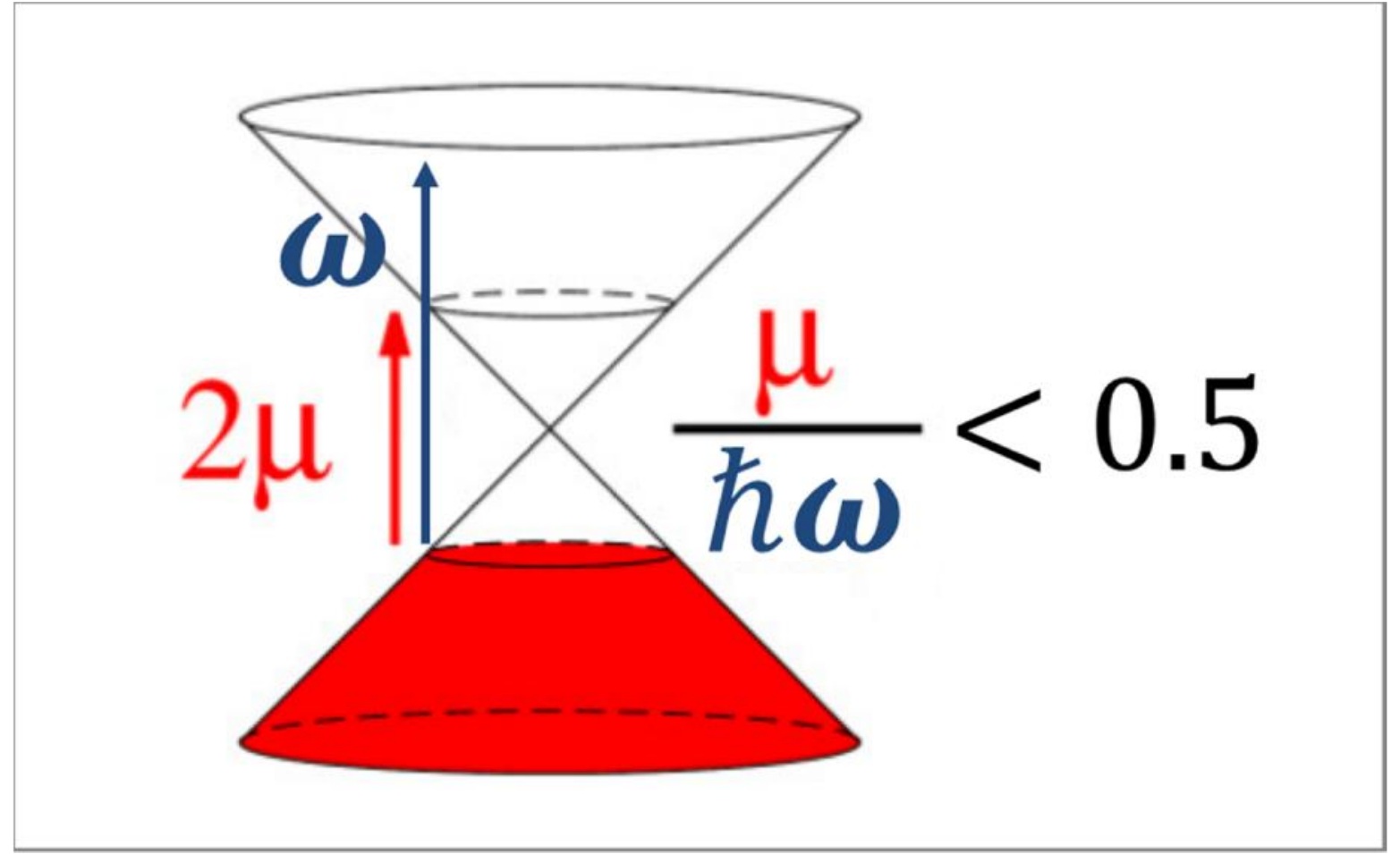}
  \caption{Band structure of doped graphene with Fermi level $\mu$. Occupied electronic states are shown in red. If the energy of an incoming photon $\hbar \omega$ exceeds $2 \mu$, it can induce a transition from the lower band to the upper one. }
 \label{Fig:InterbandTransition}
 \end{figure}

The conductivity of graphene has two physically distinct components. The first term on the right, proportional to $\mu$, corresponds to that of a free-electron gas (i.e., a Drude metal) and describes the response of carriers within a single band of graphene. The second term (in brackets) describes the effect of optically-induced transitions between the different bands of graphene. It consists of a real term (characterizing absorption) proportional to a step function, which turns on for frequencies $\omega>2\mu$, due to the availability of electron-hole transitions at these frequencies (see Fig.~\ref{Fig:InterbandTransition}). The imaginary term describes dispersive effects associated with a step-function absorption profile, as required by Kramers-Kronig relations.
In the regime $\omega\lesssim 2\mu$, interband effects can be neglected and graphene behaves like a Drude metal. In this case, graphene supports guided surface plasmon modes, like any thin conducting film~\cite{Jablan2009}. The plasmonic wavelength-frequency dispersion relation is given by $\lambda_{sp}/\lambda_0=2\alpha (\mu/\omega)$, where $\alpha$ is the fine-structure constant and $\lambda_0=2\pi c/\omega$ is the free-space wavelength~\cite{Wunsch2006}. An emitter within a distance $d\approx\lambda_{sp}$ of the surface experiences strong spontaneous emission into the plasmon modes.
\section{Linearization of the Casimir coupling Hamiltonian}\label{Sec:Linearization}
In the following, we explain how the Casimir coupling Hamiltonian can be linearized in the the weak driving limit and outline how the effective light-membrane interaction Hamiltonian (Eq.~(3) in the main text) is obtained.\\
\\The dynamics of the membrane and the emitter is governed by
\begin{eqnarray*}
\dot{\rho}&=&-i[H,\rho]+\Gamma\ \!\mathcal{D}_{{\sigma}_-}(\rho),\\
H&=&\frac{{\Delta}}{2}{\sigma}_z+\frac{\Omega}{2}{\sigma}_x+\omega_{\text{M}}{a}_{\text{M}}^{\dag}{a}_{\text{M}}+ g{x}_{\text{M}}{\sigma}_z,\\
\mathcal{D}_{\sigma_-}(\rho)&=&{\sigma}_{-}\rho{\sigma}_{+}-\frac{1}{2}{\sigma}_+{\sigma}_-\rho-\frac{1}{2}\rho{\sigma}_+{\sigma}_-,
\end{eqnarray*}
and leads to modified mechanical and emitter properties such as a $\langle \sigma_z\rangle$-dependent displacement of the steady state value of the mechanical position $\langle x_{\text{M}}\rangle_{\infty}$ and a renormalized detuning $\Delta \rightarrow\Delta +g \langle x_{\text{M}}\rangle_{\infty}$ . The differential equation above can be decomposed into an entangling part given by the coupling Hamiltonian $H_{\text{int}}= g{x}_{\text{M}}{\sigma}_z$, which creates correlations between the emitter and the light field, and a separable part
\begin{eqnarray*}
\mathcal{L}_0(\rho)&=&-i\left[\frac{{\Delta}}{2}{\sigma}_z+\frac{\Omega}{2}{\sigma}_x,\rho\right]+\Gamma\mathcal{D}_{{\sigma}_-}(\rho).
\end{eqnarray*}
We consider the case $\Gamma\gg g$, where the two level system reaches its steady state on a time scale which is fast compared to the timescale on which the exchange of information between the emitter and the membrane take place. The steady state of $\mathcal{L}_0(\rho)$,
\begin{eqnarray*}
\rho_{0}\!\!\!&=&\!\!\!\! \left(\!\!
              \begin{array}{cc}
               \! \frac{\frac{\Omega^2}{4}}{\frac{\Omega^2}{2}+{\Delta}^2+\frac{\Gamma^2}{4}}\! &\! \frac{\Omega}{2} \frac{\langle\sigma_{z}\rangle_\infty}{{\Delta}+i\frac{\Gamma}{2}}\! \\
                \!\frac{\Omega}{2} \frac{\langle\sigma_{z}\rangle_\infty}{{\Delta}-i\frac{\Gamma}{2}}\! & \! \frac{\frac{\Omega^2}{4}+{\Delta}^2+\frac{\Gamma^2}{4}}{\frac{\Omega^2}{2}+{\Delta}^2+\frac{\Gamma^2}{4}}\!  \\
              \end{array}
            \!\!\right),
\end{eqnarray*}
with
\begin{eqnarray*}
\langle\sigma_z\rangle_\infty = -1+\frac{\epsilon}{2}\!+\!\mathcal{O}(\epsilon^2),
\end{eqnarray*}
is a pure state in the weak driving limit, i.e. up to $\mathcal{O}(\epsilon)$, where $\epsilon=\frac{\Omega^2}{{\Delta}^2+(\Gamma/2)^2}$.
This allows us to introduce a unitary transformation $R$, which rotates the ground state of the emitter into the steady state of $\mathcal{L}_0$, $R^{\dag}| g\rangle \langle g|R=\rho_0$. We are interested in deviations of relevant observables from their steady state mean value and describe the interaction between the emitter and the membrane therefore in a rotated and displaced picture where
\begin{eqnarray*}
H_{\text{int}}&=&- g\left(1\!-\!\frac{\epsilon}{8}\right)\frac{\Omega}{{\Delta}^2\!+\!\frac{\Gamma^2}{4}}\left({\Delta}{\sigma}_{x}\!+\!\frac{\Gamma}{2}{\sigma}_{y}\right){x}_{\text{M}}+\mathcal{O}(\epsilon^2).
\end{eqnarray*}
Since the emitter explores only a small region on the surface of the Bloch sphere around $\rho_0$, this region can be approximated by a plane and the two level system can be treated as harmonic oscillator with quadratures
\begin{eqnarray*}
{x}_{\text{E}}&=&\frac{\bar{\alpha} \sigma_y+\bar{\beta} \sigma_x}{\sqrt{2\epsilon|\langle \sigma_z\rangle_\infty|}}, \ \ \ \  {p}_{\text{E}}=\frac{-\bar{\beta} \sigma_y+\bar{\alpha} \sigma_x}{\sqrt{2\epsilon|\langle \sigma_z\rangle_\infty|}},
\end{eqnarray*}
such that
\begin{eqnarray}\label{Eq:HamiltonianME}
H_{\text{int}}&=& \bar{g}\sqrt{\epsilon}{x}_{\text{M}}x_{\text{E}} ,
\end{eqnarray}
to first order in $\epsilon$, where
\begin{eqnarray}\label{Eq.AlphaBeta}
\bar{g}=-g\sqrt{2}\left(\!1\!-\!\frac{3}{8}\epsilon\right), \ \ \ \bar{\alpha}=\frac{\Omega \frac{\Gamma}{2}}{{\Delta}^2 \!+\!\frac{\Gamma^2}{4}},\ \ \
\bar{\beta}=\frac{\Omega {\Delta}}{{\Delta}^2 \!+\!\frac{\Gamma^2}{4}}.
\end{eqnarray}
The emitter interacts with the membrane through Eq.~(\ref{Eq:HamiltonianME}) and with the light-field via the standard optical Bloch equations. For $\epsilon\ll1$, the emitter can be adiabatically eliminated, which yields an effective interaction between the membrane and the light field
\begin{eqnarray}\label{Eq:EffectiveHamiltonian}
H_{\text{\tiny{ML}}}=\kappa\ \!{x}_{\text{M}} p_{\text{L}},
\end{eqnarray}
where $\kappa=\bar{g}\sqrt{\epsilon\Gamma_{\text{det}}}/\Gamma$ and $p_{\text{L}}$ is the light field quadrature, which couples to $x_E$ (see Sec.~\ref{Sec:Protocol}). A detailed derivation of the corresponding equations of motion is provided in Sec.~\ref{Sec:TimeEvolution}.
Eq.~(\ref{Eq:EffectiveHamiltonian}) describes the mapping of displacements of the membrane $x_{\text{M}}$ to phase shifts on the scattered light field that are described by the quadratures $x_{L}$ and $p_L$ (see below). These phase shift can be very efficiently measured against a reference beam using homodyne detection~\cite{Loudon}.
\section{Read-out scheme}\label{Sec:Protocol}
This section is devoted to the read-out of the Casimir potential induced level shifts using coherent light fields.
In Sec.~\ref{Sec:TimeEvolution}, the effective emitter-mediated time evolution of the membrane and the light field is derived and in Sec.~\ref{Sec:MechanicalVariance} we explain how the conditional variance of the membrane can be calculated if the scattered light field is measured by homodyne detection.
Throughout this part of the Supplemental Material, we will use dimensionless mechanical quadratures $x_{\text{m}}=x_{\text{M}}\ \!\sqrt{\omega_{\text{M}} m}$, $p_{\text{m}}=p_{\text{M}}\ \!\sqrt{1/(\omega_{\text{M}} m)}$ (as above, we use $\hbar=1$). With this notation, the Casimir coupling Hamiltonian derived in Sec.~\ref{Sec:Linearization} is given by
\begin{eqnarray}\label{Eq:SI:CasimirH}
H_{\text{int}}=\bar{g}_{\text{m}}\sqrt{\epsilon}\ \!x_{\text{E}}x_{\text{m}}, \  \  \ \bar{g}_{\text{m}}=\bar{g}\ \!(m\omega_{\text{M}})^{-\frac{1}{2}}.
\end{eqnarray}
\subsection{Time evolution of the emitter, the membrane and the light field}\label{Sec:TimeEvolution}
In this section, we derive the input-output relations for the light field and the membrane in the weak driving limit by adiabatically eliminating the emitter.
\subsubsection{Light-emitter interaction and adiabatic elimination of the excited state }
In the following, we consider the evolution of the light field and the emitter. As explained above, the properties of the emitter are read out by applying a laser beam and detecting the phase shift that has been acquired by the light field. The phase shift on the light field are here described in terms of the light field quadratures $x_{\text{L}}$ and $p_{\text{L}}$. They describe the in-phase and out-of-phase component of the light with respect to some reference laser field~\cite{Loudon}. The former corresponds to the sine component (with a phase difference of $\phi=0$ with respect to the reference beam) of the electromagnetic field. The latter corresponds to the cosine component (with a phase difference of $\phi=\pi/2$).\\
We use here spatially localized light modes $x_L(t)$, $p_L(t)$ with commutation relations $[x_L(t),p_L(t')]=i\delta(t-t')$. The light mode corresponding to $x_L(t)$, $p_L(t)$ interacts with the emitter at time $t$ through the dipole interaction, resulting in the transformation
\begin{eqnarray}\label{Eq:LightFieldBasic}
\left(
  \begin{array}{c}
    {x}^{\text{out}}_{\text{L}}(t) \\
    {p}^{\text{out}}_{\text{L}}(t) \\
  \end{array}
\right)\!\!&=&\!\!\left(
  \begin{array}{c}
    \!{x}^{\text{in}}_{\text{L}}(t)\! \\
    \!{p}^{\text{in}}_{\text{L}}(t)\! \\
  \end{array}
\right)\!+\!\sqrt{\Gamma_{\text{det}}}\left(
                                    \begin{array}{c}
                                    \!\!  -p_{\text{E}}(t) \!\!\\
                                    \!\!  x_{\text{E}}(t) \!\!\\
                                    \end{array}
                                  \right),
\end{eqnarray}
where the superscripts "in" and "out" label the variables before and after the interaction.
The emitter is subject to three different types of interactions. It couples to the light field and interacts with the membrane through the Casimir potential, as described by  Eq.~(\ref{Eq:SI:CasimirH}). Moreover, the emitter can scatter light into channels which are not measured. The latter is taken into account by introducing noise modes ${x}_{\text{N}}(t)$, ${p}_{\text{N}}(t)$ with $[{x}_{\text{N}}(t),{p}_{\text{N}}(t')]=i\delta(t-t')$ such that
\begin{eqnarray*}
\left(
  \begin{array}{c}
    \dot{x}_{\text{E}}(t) \\
    \dot{p}_{\text{E}}(t) \\
  \end{array}
\right)
&=&\sqrt{\Gamma_{\text{det}}}\left(
                                          \begin{array}{c}
                                            -{p}^{\text{in}}_{\text{L}}(t) \\
                                            {x}^{\text{in}}_{\text{L}}(t) \\
                                          \end{array}
                                        \right)
+\sqrt{\Gamma_{\text{N}}}\left(
                                          \begin{array}{c}
                                            -{p}^{\text{in}}_{\text{N}}(t)\! \\
                                            {x}^{\text{in}}_{\text{N}}(t)\! \\
                                          \end{array}
                                        \right)
-\bar{g}_{\text{m}}\sqrt{\epsilon}\left(
                                                                             \begin{array}{c}
                                                                               0 \\
                                                                               x_{\text{m}}(t) \\
                                                                             \end{array}
                                                                           \right)-\frac{\Gamma}{2}
\left(
  \begin{array}{c}
    x_{\text{E}}(t) \\
    p_{\text{E}}(t) \\
  \end{array}
\right),
\end{eqnarray*}
where $\Gamma=\Gamma_{\text{det}}+\Gamma_{\text{N}}$. In the weak driving limit, where $\epsilon=\frac{\Omega^2}{\Delta^2+\frac{\Gamma^2}{4}}\ll 1$, the population of the excited state is negligible and the emitter can be adiabatically eliminated. For $\epsilon \ll 1$ and $\omega_{\text{M}}\ll \Gamma$,
\begin{eqnarray}\label{Eq:AdiabaticEliminationAtom}
\left(
  \begin{array}{c}
    x_{\text{E}}(t) \\
    p_{\text{E}}(t) \\
  \end{array}
\right)&=&\frac{2\sqrt{\Gamma_{\text{det}}}}{\Gamma}\left(
                                          \begin{array}{c}
                                            -{p}_{\text{L}}^{\text{in}}(t) \\
                                            {x}_{\text{L}}^{\text{in}}(t) \\
                                          \end{array}
                                        \right)+\frac{2\sqrt{\Gamma_{\text{N}}}}{\Gamma}\left(
                                          \begin{array}{c}
                                            -{p}_{\text{N}}^{\text{in}}(t) \\
                                            {x}_{\text{N}}^{\text{in}}(t) \\
                                          \end{array}
                                        \right)
-\frac{2\bar{g}_{\text{m}}\sqrt{\epsilon}}{\Gamma}\left(
                                                                     \begin{array}{c}
                                                                       0 \\
                                                                       x_{\text{m}}(t) \\
                                                                     \end{array}
                                                                   \right).
\end{eqnarray}
By inserting this expression into the evolution equation for the membrane and the light field, effective input-output relations for the mechanical and the photonic system can be obtained, which do not include the emitter any more.
\subsubsection{Effective time evolution of the membrane and the light field}
The membrane evolves under the Hamiltonian $H_{\text{\tiny{membrane}}}=\frac{\omega_{\text{M}}}{2}(x_{\text{m}}^2+p_{\text{m}}^2)+\bar{g}_{\text{m}}\sqrt{\epsilon}\ \!x_{\text{E}}x_{\text{m}}$ and is subject to mechanical damping.  We consider here two different damping models and analyze the case of symmetric damping when position and momentum are damped with equal rates $\gamma_x=\gamma_p=\gamma/2$, and pure momentum damping $\gamma_x=0$, $\gamma_p=\gamma$.
In the case of symmetric damping, the time evolution of the membrane is given by
\begin{eqnarray}\label{Eq:1}
\left(
  \begin{array}{c}
    \dot{x}_{\text{m}}(t) \\
    \dot{p}_{\text{m}}(t) \\
  \end{array}
\right)_{\text{\tiny{sym}}}&=&\left(
            \begin{array}{cc}
              -\frac{\gamma}{2} & \omega_{\text{M}} \\
              - \omega_{\text{M}} & -\frac{\gamma}{2} \\
            \end{array}
          \right)
\left(
  \begin{array}{c}
    x_{\text{m}}(t) \\
    p_{\text{m}}(t) \\
  \end{array}
\right)+\sqrt{\gamma}\left(
                                     \begin{array}{c}
                                        f^{\text{in}}_{x,\text{\tiny{sym}}}(t)\\
                                        f^{\text{in}}_{p,\text{\tiny{sym}}}(t)\\
                                     \end{array}
                                   \right)
-\bar{g}_{\text{m}}\sqrt{\epsilon}\left(
                                                                            \begin{array}{c}
                                                                              0 \\
                                                                              x_{\text{E}}(t) \\
                                                                            \end{array}
                                                                          \right),
\end{eqnarray}
where $\gamma$ is the mechanical decay rate and $f^{\text{in}}_{x,\text{\tiny{sym}}}$, $f^{\text{in}}_{p,\text{\tiny{sym}}}$ are the associated Langevin noise operators with $[f^{\text{in}}_{x,\text{\tiny{sym}}}(t),f^{\text{in}}_{p,\text{\tiny{sym}}}(t')]=i\delta(t-t')$. The noise correlation functions are given by $\langle f^{\text{in}}_{x,\text{\tiny{sym}}}(t)f^{\text{in}}_{x,\text{\tiny{sym}}}(t') \rangle=\langle f^{\text{in}}_{p,\text{\tiny{sym}}}(t)f^{\text{in}}_{p,\text{\tiny{sym}}}(t') \rangle=\delta(t-t')(2n_{\text{th}}+1)$, where $n_{\text{th}}$ is the thermal occupation number and is given by $n_{\text{th}}=k_{\text{B}}T_{\text{bath}}/\omega_{\text{m}}$. $k_{\text{B}}$ is the Boltzmann constant and $T_{\text{bath}}$ is the temperature of the bath of the membrane.\\

In the following, we derive the effective evolution for pure momentum damping. The symmetric case can be treated in an analogous fashion.
Physically, many known damping mechanisms lead to momentum- rather than position damping. However, the quantized description of pure momentum damping is a complicated problem which is for example addressed in the Caldeira-Leggett model~\cite{CaldeiraLeggett,Banerjee2004} and involves non-Markovian noise operators. We use here a simplified Markovian model, which can be understood as the quantum analogue of classical Brownian motion~\cite{GardinerZoller,Breuer}. A direct Markovian quantum analogue of the equations describing classical Brownian motion does in general not preserve the positivity of the density matrix describing the quantum state~\cite{Haake85}. This can be corrected by adding an appropriate noise term in the evolution of $x_{\text{M}}(t)$~\cite{Damping1,Damping2}. The corresponding quantum Langevin equations for the mechanical quadratures are given by
\begin{eqnarray*}
\left(
  \begin{array}{c}
    \dot{x}_{\text{m}}(t) \\
    \dot{p}_{\text{m}}(t) \\
  \end{array}
\right)\!\!&=&\!\!\left(
            \begin{array}{cc}
                0\! &\! \omega_{\text{M}} \\
              - \omega_{\text{M}} & -\gamma \\
            \end{array}
          \right)
\left(
  \begin{array}{c}
    x_{\text{m}}(t) \\
    p_{\text{m}}(t) \\
  \end{array}
\right)+\sqrt{\gamma}\left(
                                     \begin{array}{c}
                                        f^{\text{in}}_{x}(t)\\
                                        f^{\text{in}}_{p}(t)\\
                                     \end{array}
                                   \right)
-\bar{g}_{\text{m}}\sqrt{\epsilon}\left(
                                                                            \begin{array}{c}
                                                                              0 \\
                                                                              x_{\text{E}}(t) \\
                                                                            \end{array}
                                                                          \right),
\end{eqnarray*}
with $[f^{\text{in}}_{x}(t),f^{\text{in}}_{p}(t')]=i\delta(t-t')$ and noise correlation functions $\langle f^{\text{in}}_{x}(t)f^{\text{in}}_{x}(t') \rangle=\delta(t-t')(2n_{\text{th}}+1)^{-1}$, $\langle f^{\text{in}}_{p}(t)f^{\text{in}}_{p}(t') \rangle=\delta(t-t')(2n_{\text{th}}+1)$~\cite{Damping1,Damping2}.\\

We consider now the time evolution of the membrane in the interaction picture with respect to the free mechanical Hamiltonian $H_M=\frac{\omega_{M}}{2}(x_m^2+p_m^2)$, i.e. in a co-rotating frame. The corresponding transformed mechanical variables are given by
\begin{eqnarray}\label{Eq:RotatingFrame}
\left(
  \begin{array}{c}
   \! \tilde{x}_{\text{m}}(t)\! \\
   \! \tilde{p}_{\text{m}}(t)\! \\
  \end{array}
\right)\!\!&=&\!\!\left(
            \begin{array}{cc}
             \!\! \cos(\omega_{\text{M}}t)\!\! & \!-\!\sin(\omega_{\text{M}}t)\!\! \\
             \!\! \sin(\omega_{\text{M}}t)\!\! & \!\!\cos(\omega_{\text{M}}t)\!\! \\
            \end{array}
          \right)
\left(
  \begin{array}{c}
   \!\! x_{\text{m}}(t)\!\! \\
   \!\! p_{\text{m}}(t)\!\! \\
  \end{array}
\right)
\end{eqnarray}
and evolve according to
\begin{eqnarray}\label{Eq:MembraneDGL}
\left(
  \begin{array}{c}
    \dot{\tilde{x}}_{\text{m}}(t) \\
    \dot{\tilde{p}}_{\text{m}}(t) \\
  \end{array}
\right)&=&\gamma_R\left(
  \begin{array}{c}
    \tilde{x}_{\text{m}}(t) \\
    \tilde{p}_{\text{m}}(t) \\
  \end{array}
\right)
+ \sqrt{\gamma}\!\left(
          \begin{array}{c}
            \cos(\omega_{\text{M}}t)f^{\text{in}}_{x}(t)-\sin(\omega_{\text{M}}t)f^{\text{in}}_{p}(t) \\
            \sin(\omega_{\text{M}}t) f^{\text{in}}_{x}(t)+\cos(\omega_{\text{M}}t) f^{\text{in}}_{p}(t)\\
          \end{array}
        \right)\\
&+&\frac{2\bar{g}_{\text{M}}\sqrt{\epsilon\ \!\Gamma_{\text{det}}}}{\Gamma}
\left(
                                \begin{array}{c}
                                  -\sin(\omega_{\text{M}}t)\\
                                  \cos(\omega_{\text{M}}t) \\
                                \end{array}
                              \right){p}^{\text{in}}_{\text{L}}(t)
+\frac{2\bar{g}_{\text{M}}\sqrt{\epsilon\ \!\Gamma_{\text{N}}}}{\Gamma}
\left(
                                \begin{array}{c}
                                  -\sin(\omega_{\text{M}}t)\\
                                  \cos(\omega_{\text{M}}t) \\
                                \end{array}
                              \right){p}^{\text{in}}_{\text{N}}(t)\nonumber,
\end{eqnarray}
where
\begin{eqnarray*}
\gamma_R&=&\gamma\left(
            \begin{array}{cc}
             - \sin^2(\omega_{\text{m}}t)&  \cos(\omega_{\text{m}}t) \sin(\omega_{\text{m}}t)  \\
             \cos(\omega_{\text{m}}t) \sin(\omega_{\text{m}}t)&- \cos^2(\omega_{\text{m}}t)  \\
            \end{array}
          \right).
\end{eqnarray*}
The equations for the evolution of the light field quadratures read
\begin{eqnarray}\label{Eq:LightFieldDGL}
\left(
  \begin{array}{c}
   \!\! {x}_{\text{L}}^{\text{out}}(t) \!\!\\
    \!\!{p}_{\text{L}}^{\text{out}}(t)\!\! \\
  \end{array}
\right)\!\!\!&=&\!\!\frac{2\bar{g}_{\text{m}}\sqrt{\epsilon\Gamma_{\text{det}}}}{\Gamma}\left(
                                                                                               \begin{array}{c}
                                                                                                 \!\!\cos(\omega_{\text{M}}t)\tilde{x}_{\text{m}} (t) \!+\!\sin(\omega_{\text{M}}t) \tilde{p}_{\text{m}}(t)\!\! \\
                                                                                                \!\! 0 \!\!\\
                                                                                               \end{array}
                                                                                             \right)                                                                                     \!+\! \left( 1\! -\! \frac{2\Gamma_{\text{det}}}{\Gamma} \right) \left(
  \begin{array}{c}
   \!{x}_{\text{L}}^{\text{in}}(t)\!\! \\
   \!{p}_{\text{L}}^{\text{in}}(t)\!\! \\
  \end{array}
\right)
-\frac{2\sqrt{\Gamma_{\text{det}}\Gamma_{\text{N}}}}{\Gamma}\!\left(
  \begin{array}{c}
   \! {x}^{\text{in}}_{\text{N}}(t)\!\!\\
    \!{p}^{\text{in}}_{\text{N}}(t)\!\! \\
  \end{array}
\right)\!\!,
\end{eqnarray}
where Eq.~(\ref{Eq:LightFieldBasic}), Eq.~(\ref{Eq:AdiabaticEliminationAtom}) and Eq.~(\ref{Eq:RotatingFrame}) have been used.
The time evolution equations for the mechanical and light-field variables Eq.~(\ref{Eq:MembraneDGL}) and Eq.~(\ref{Eq:LightFieldDGL}) correspond to an effective interaction between the membrane and light,
$H_{\text{ML}}=\kappa x_{\text{M}}p_{\text{L}}(t)$
with effective coupling rate
$\kappa=2\bar{g}\sqrt{\epsilon \Gamma_{\text{det}}}/\Gamma$.
\subsection{Calculation of the conditional variance}\label{Sec:MechanicalVariance}
Eq.~(\ref{Eq:LightFieldDGL}) shows that the mechanical position is mapped to the $x$-quadrature of the light field and can accordingly be read-out by monitoring $x_{\text{L}}$. As discussed in the following, continuous measurements of $x_{\text{L}}$ lead to a reduced conditional variance of the position of the membrane.\\
The term conditional variance refers to the variance that is obtained if the measurement results are known. The term unconditional variance describes the case where the light field is not measured or if the measurement results are not taken into account. In the setting considered here, the conditional variance of the atomic position $V_x$ can be reduced below $x_{\text{ZPM}}^2$, while the unconditional state does not exhibit squeezing. This is due to the fact that the measurements on the light field yield probabilistic outcomes which result in random displacements of the mechanical state in phase space.\\

In the following, we explain how the conditional variance can be calculated~\cite{Muschik2012}. For convenience, we discretize time using infinitesimally short time intervals of duration $\tau\ll \omega_{\text{M}}^{-1}$, $\kappa^{-2}$, $(\gamma\cdot n_{\text{th}})^{-1}$ and discretized light modes $x_{L,n}^{in}=\sqrt{\tau}x_{L}^{in}(n\tau)$, $p_{L,n}^{in}=\sqrt{\tau}p_{L}^{in}(n\tau)$.\\
The light-membrane interaction discussed above leads to an entangled state between the membrane and the light. As outlined above, measurement on the latter allow one to infer information of the former such that a squeezed state is generated. More specifically, for each time step, a light mode in vacuum $|0\rangle_n$ couples to the membrane in state $|\Psi_{\text{M}}(n\tau)\rangle$ through the interaction Hamiltonian $H_{\text{ML}}$. The subsequent measurement yields outcome $o_n$ with probability $p_n$. In this case, we obtain the conditional state of the membrane $|\Psi_{\text{M}}([n+1]\tau)\rangle=\frac{1}{\sqrt{p_n}} _n\!\langle o_n|e^{-iH_{\text{ML}}\tau}|0\rangle_n |\Psi_M(n \tau)\rangle$ and the corresponding unconditional state is given by $\rho([n+1]\tau)=\sum_{n} M_n \rho(n\tau)M_{n}^{\dag}$, where $M_n=\!_n\langle o_n|e^{-iH_{\text{ML}}\tau}|0\rangle_n $.\\

This process can be conveniently described in terms of covariance matrices using the Gaussian formalism~\cite{Giedke2002,GaussianReview}. The covariance matrix of a continuous variable system with $m$ modes that are each described by the quadratures $x$ and $p$ is given by
$\Gamma_{ij}=\langle\{\langle R_{i}-\langle R_{i}\rangle,R_{j}-\langle R_j \rangle\rangle\}_+ \rangle$, where $\{\cdot,\cdot\}$ is the anticommutator and  $\mathbf{R}=(x_1,p_1,...x_m,p_m)^T$. The covariance matrix of a thermal state is for example given by $\Gamma_{\text{th}}=(2 n_{th}+1)\cdot\openone$.
Unitary time evolutions $\mathbf{R}(t)=S(t) \mathbf{R}^{\text{in}}$ can be parametrized by a time evolution matrix $S$ such that $\Gamma(t)=S(t) \Gamma^{\text{in}}S(t)^T $. Using this notation, the time evolution of the membrane and the light field given by Eq.~(\ref{Eq:LightFieldDGL}) and Eq.~(\ref{Eq:MembraneDGL}) can be cast in the form $\Gamma([n+1]\tau)=S(n\tau)\Gamma(n\tau)S^{T}(n\tau)$.
$\Gamma([n+1]\tau)$ is here a $8\times8$ square matrix corresponding to $\mathbf{R}(n\tau)=(x_{\text{m}}(n\tau),p_{\text{m}}(n\tau),x_{\text{L,}n},p_{\text{L,}n},x_{\text{N,}n},p_{\text{N,}n},f_{x,n},f_{p,n})^T$.
The update of the mechanical state through the measurement of the light~\cite{Giedke2002} can be calculated by considering the $4\times4$ block of this matrix $\Gamma_{\text{ML}}$ that corresponds to the mechanical and photonic modes
\begin{eqnarray*}
\Gamma_{\text{ML}}=\left(
                    \begin{array}{cc}
                      \Gamma_{\text{M}} & \Gamma_{\text{coh}} \\
                       \Gamma^T_{\text{coh}} & \Gamma_{\text{L}} \\
                    \end{array}
                  \right),
\end{eqnarray*}
and using the formula
\begin{eqnarray}\label{Eq:MeasurementUpdate}
\Gamma_{\text{M}}'=\Gamma_{M}-\Gamma_{\text{\tiny{coh}}}(\Gamma_{\text{L}}+\tilde{\gamma}_{L})^{-1}\Gamma_{\text{\tiny{coh}}}^T.
\end{eqnarray}
$\Gamma_{\text{M}}'$ is the updated $2\times2$ matrix, which describes the conditional mechanical state after the measurement, and
\begin{eqnarray*}
\tilde{\gamma}_{L}=\left(
                     \begin{array}{cc}
                       r^{-1} & 0 \\
                       0 & r \\
                     \end{array}
                   \right)
\end{eqnarray*}
is the covariance matrix of the state onto which the photonic mode is projected. A perfect measurement of $x_{\text{L}} $corresponds to $r\rightarrow \infty$.

For example, if the initial state of the mechanical system is a thermal state and the dynamics is solely governed by the interaction Hamiltonian $H_{ML}$ (which is the case in the short time limit for perfect detection), Eq.~(\ref{Eq:MeasurementUpdate}) yields directly
\begin{eqnarray*}
\dot{V_x}(t)&=&-\kappa^2 V_x^2(t),\ \ \ \  \dot{V_p}(t)=\kappa^2,
\end{eqnarray*}
such that
\begin{eqnarray*}
V_x(t)=\frac{1}{(V_x^{in})^{-1}+\kappa^2 t}, \ \ \ V_p(t)=V_p^{\text{in}}+\kappa^2 t.
\end{eqnarray*}
This underlying mechanism which leads to a squeezing in the mechanical $x$-quadrature and an antisqueezing in $p_{\text{m}}$ is complicated by the effects of imperfect detection, the coupling to a thermal bath and the rotation in phase space~\cite{Hammerer2005}. We consider here the measurement process in the rotating frame, since the co-rotating variables $\tilde{x}_{\text{m}}$ and $\tilde{p}_{\text{m}}$ are the relevant observables that can be accessed typically. Since the interaction Hamiltonian facilitates a mapping of $x_{\text{m}}$ onto $x_{\text{L}}$, $\tilde{x}_{\text{m}}$ and $\tilde{p}_{\text{m}}$ are mapped and squeezed alternatingly at a frequency $\omega_{\text{M}}$, which gives rise to the oscillations in Fig.~4b in the main text. In the non-rotating frame, the conditional variance of $x_{\text{m}}$ decreases quickly during a short time interval and reaches then a stationary value with constant squeezing.
%
%

%
\end{document}